\documentclass[12pt,a4paper]{article}
\begin{document}
\newcommand{\eq}[1]{~(\ref{#1})}
\newcommand{\BEQ}{\begin{equation}}
\newcommand{\EEQ}{\end{equation}}
\newcommand{\norm}[1]{\mbox{$\left| \left| #1 \right| \right| $}}
\newcommand{\abs}[1]{\mbox{$\left| #1 \right| $}} 
\newcommand{\bra}[1]{\mbox{$\langle \left. #1 \right| $}}
 \newcommand{\ket}[1]{\mbox{$\left|  #1 \rangle \right. $}}
\newcommand{\DEG}[1]{\mbox{$ #1^{\rm o}$}}
\newcommand{\lappr}{\mbox{$\stackrel{<}{\sim}$}} 
\newcommand{\gappr}{\mbox{$\stackrel{>}{\sim}$}} 
\newcommand{\mr}[1]{\mbox{\rm #1}} 
\newcommand{\ETC}{\mbox{\em etc.\/ }}
\newcommand{\VIZ}{\mbox{\em viz.\/ }}
\newcommand{\CF}{\mbox{\em cf.\/ }}
\newcommand{\IE}{\mbox{\em i.e. \/}}
\newcommand{\ETAL}{\mbox{\em et. al.\/ }}
\newcommand{\EG}{\mbox{\em e.g.\/ }}
%
%
\begin{flushright}
DFF-269/02/1997 (Florence)\\
JHU--TIPAC--97002 (Johns Hopkins)\\
February 1997 
\end{flushright}
\vspace*{8mm}
\begin{center}
{\Large\bf Neutrino Flavor Conversion in Random Magnetic Fields}\\[5truemm]
G. Domokos and S. Kovesi--Domokos\\[2mm]
Dipartimento di Fisica, Universit\'{a} di Firenze\\
Florence, Italy\\[1mm]
and\\[1mm]
The Henry A. Rowland Department of Physics and Astronomy\\
The Johns Hopkins University\footnote{Permanent address. 
E--mail:~SKD@HAAR.PHA.JHU.EDU}\\
Baltimore, MD 21218
\end{center}
\vspace*{4mm}
\begin{quote}
If massive neutrinos possess magnetic moments,
a magnetic field can cause a spin flip. In the case of Dirac neutrinos
the spin flip converts an active neutrino into a sterile one and {\em
vice versa\/}. By contrast,
if neutrinos are Majorana particles, a spin flip converts them to a neutrino
of a different flavor. We examine the behavior of neutrinos in a random
 magnetic field as it occurs, for instance, in certain astronomical
objects, such as an active galactic nucleus. Both Dirac and Majorana 
neutrinos behave ergodically:
independently of their initial density matrix, they tend towards an 
equipartition
of the helicity states. As a result, about half of the Dirac neutrinos
produced becomes sterile. For Majorana neutrinos, there will be an approximate
equipartition of flavors, independently of the production mechanism.
\end{quote}
\vspace*{5mm}
\begin{flushleft}
\end{flushleft}\vspace{3mm}
According to the standard model of electroweak interactions, 
neutrinos cannot possess magnetic moments.
The possibility of giving magnetic moments to neutrinos arises in minimal
extensions of the standard model: once right handed neutrinos are
introduced, chirality violating,
Pauli type magnetic interactions with an electromagnetic field can
 be introduced.
There is an important difference between the physical effects caused
by such interactions, depending on whether neutrinos are Dirac or
Majorana particles. The Dirac case is  somewhat uninteresting:
a chirality flipping interaction causes a conversion between  
active and sterile neutrino species. By contrast, in the case of Majorana
neutrinos, different chiralities must correspond to different flavors
and thus, a chirality flipping transition causes transitions between
flavors at the same time. The minimally extended standard model gives
rise -- {\em via} higher order loop effects -- to anomalous magnetic
moments; however, their magnitudes are rather small and one can
hardly hope to ever detect them. One  needs truly ``new physics''
beyond the standard model in order to generate detectable magnetic
moments of neutrinos. There exist rather strong -- albeit somewhat
model dependent -- upper limits of flavor diagonal magnetic moments,
typically in the range of $10^{-12}\mu_{B}$, where
$\mu_{B}$ is the Bohr magneton. The upper limits for
transition moments are generally weaker, typically 
by some three to four orders
of magnitude; see, for instance, ref.~\cite{kimpevsner} for a
recent review. Thus, the experimental identification of any effect caused by
the presence of  a magnetic interaction of neutrinos is of utmost
interest, for it provides a handle on physics beyond the standard model.

In this work we examine the fate of neutrinos possessing 
anomalous magnetic moments in random 
magnetic fields. Such fields may serve as models of the chaotic fields expected
to exist within certain astrophysical objects, in particular, within 
active galactic
nuclei (AGN). An AGN is a particularly interesting environment, since,
according to many speculations, it is a source of ultra high energy neutrinos,
see, for instance the proceedings of the Hawaii meeting~\cite{hawaii}.

It is generally assumed that most neutrinos emerge from the production and
subsequent decay of hadrons. Assuming that one can
extrapolate accelerator based data to the relevant (PeV or higher) energies,
the overwhelming majority of hadrons produced in a hadron collision consists 
of pions. Hence, the charged leptons produced in a neutrino telescope
observing the point source 
must be mostly muons\footnote{In a previous work~\cite{resonant} we 
discussed the possibility that resonant interactions in the hot plasma
present in an AGN  may cause the appearance of different 
flavors of neutrinos. Also,
Learned and Pakvasa~\cite{learnedpakvasa} considered
flavor conversion due to long range oscillations.}. We argue that,
under appropriate circumstances, a random magnetic field equilibrates
the helicities; hence, in he case of Majorana neutrinos, the flavors as well.
In this work, we consider the behavior of a single spin-1/2 field,
without paying detailed attention to the flavor structure.

In order to describe the average behavior of a neutrino in a random
magnetic field, one has to solve the dynamical equations governing
the propagation in an arbitrary magnetic field. The solution then has
to be averaged over the ensemble of  magnetic fields. 

We use the front form of
dynamics~\cite{dirac}. As explained in a previous 
work~\cite{heneutr}, this
formulation of dynamics is advantageous in a situation in which one
considers the propagation of high energy particles ($E\gg m$, where
$m$ is the rest mass) and in which certain discrete symmetries, such as
$C$ and $P$ play no significant role. Clearly, the propagation of high
energy neutrinos falls into this category.

We begin with the usual Dirac Lagrangian of a particle in an external 
electromagnetic field, $F_{\mu \nu}$:
\BEQ \label{diraclagrangian}
L = \overline{\psi}\left(i \gamma^{\mu}\partial_{\mu} + m 
+ \frac{1}{2}\mu F^{\mu \nu}\sigma_{\mu \nu}\right)\psi
\EEQ
We work in the rest frame of the magnetic field.  Assuming the 
field to be a static one\footnote{In physical terms,
this means that the characteristic time scale of change of  the field is
large compared to the time of passage of neutrinos.}, we can set $F_{0i}=0$, $F_{ij}=\epsilon_{ijk}B_{k}$.
In the case of 
interest one has to solve the Dirac equation in an {\em arbitrary}
static magnetic field, since we want to average the solution over an
ensemble of  the $B_{i}$. No explicit solution is 
known for such a problem. However, we proceed to show that in the
{\em high energy limit} the problem can be solved in a closed form.

We introduce a coordinate system in which two of the coordinates
are null directions corresponding to characteristic lines of a relativistic 
wave equation, \VIZ:
\BEQ
t=\frac{1}{\sqrt{2}}\left( x^{0} - x^{3}\right), z=\frac{1}{\sqrt{2}}\left(
x^{0} + x^{3}\right) \quad {\rm and} \quad x^{A}; \quad (A=1,2).
\label{nullcoordinates}
\EEQ
Correspondingly, the metric is of the form,
\BEQ
g_{zt}=g_{tz}=1, \quad g_{AB}= -\delta_{AB},
\label{metric}
\EEQ
and all other components vanish.

A Dirac spinor can be decomposed along the null directions 
given in \eq{nullcoordinates} by introducing
the mutually orthogonal projectors,
\BEQ P_{t}=\frac{1}{2}\gamma_{t}\gamma^{t}, \quad P_{z}=\gamma_{z}\gamma^{z}
\label{nullprojectors}
\EEQ
In what follows, we use the shorthand,
\BEQ
\phi = P_{t}\psi, \quad \chi = P_{z}\psi
\label{shortnames}
\EEQ

It is a straightforward matter to decompose \eq{diraclagrangian}
according to the conjugate null directions and express it in terms of
the variables $\phi$ and $\chi$. The purpose of such an exercise is a very 
simple one. If, for the sake of definiteness, $t$ is regarded the ''time''
variable describing the dynamics of the system, only $\phi$ obeys
an equation containing $\partial_{t}$. Hence, the component of the 
Dirac spinor
corresponding to the conjugate null direction
 obeys only an equation of constraint. The constraint 
can be, in turn, solved before one attempts to attack the problem of dynamics.

After carrying out the decomposition of \eq{diraclagrangian} according
to the null directions, one finds: 
\begin{eqnarray}
L &=& \sqrt{2}\left[ \phi^{\dag}\left( i \partial_{t} -i \sqrt{2} \mu
\epsilon^{AB}\gamma_{A}B_{B}\right) \phi \right. \nonumber \\
  &+& \left. \chi^{\dag}\left( i\partial_{z} - i\sqrt{2} \mu
\epsilon^{AB} \gamma_{A}B_{B}\right) \chi \right] \nonumber \\
  &+& \frac{1}{\sqrt{2}}\left[ \phi^{\dag}\gamma^{z}\left(
i\gamma^{A}\partial_{A} +m - \frac{i}{\sqrt{2}}\mu B_{3}\,\epsilon_{AB}
\gamma ^{A} \gamma^{B}\right)\chi \right. \nonumber \\
  &+& \left.  \chi^{\dag}\gamma^{t}\left( i \gamma^{A}\partial_{A} + m 
 - \frac{i}{\sqrt{2}}\mu B_{3}\,\epsilon_{AB}
\gamma^{A}\gamma^{B}\right)\phi \right]
\label{decomposedlagrangian}
\end{eqnarray}

Variation of \eq{decomposedlagrangian} with respect to 
$\chi^{\dag}$ gives the constraint. The constraint can be solved in a
straight forward fashion and eliminated from the Lagrangian. The
result is conveniently written in Hamiltonian form:
\begin{eqnarray}
L & = & \pi \partial_{t}\phi -H \nonumber \\ 
H & = & -2\mu \phi^{\dag}\sigma^{A}B_{A}\phi \nonumber \\
  & + & \phi^{\dag}\left( -i \sigma_{B}\epsilon^{BC}p_{C} +
m -\mu \sqrt{2}B_{3}\sigma_{3}\right)\nonumber \\ 
& \times & \Omega  \left( -B^{A}\right)\nonumber \\
 & \times & \left( i\sigma_{R}\epsilon^{RS}p_{S}
 + m -\mu \sqrt{2} B_{3}\sigma{3}\right)\phi\nonumber \\
  &   &
\label{Hamiltonian}
\end{eqnarray}
Solving the constraint eliminates two components of
the original, four component Dirac spinor. Therefore, instead of the
original Dirac matrices one can use $2\times 2$ Pauli matrices.
One easily verifies that $-i\epsilon^{AB}\gamma_{B}\rightarrow
\sigma^{A}$ gives the correct representation. We also introduced the
Hermitean operator, $p_{A} = -i \partial_{A}$ for the transverse
degrees of freedom.

The canonical momentum is given by $\pi = i \sqrt{2}\, \phi^{\dag}$.
(Of course, the odd looking factor of $\sqrt{2}$ in the 
definition of the canonical momentum can be eliminated by
rescaling the time variable.)
In equation \eq{Hamiltonian}, $\Omega$ is an operator with matrix 
elements:
\BEQ
\bra{ z} \Omega \left( B^{A}\right) \ket{ z'} 
 = \frac{i}{\sqrt{2}}\exp\left(\mu \sqrt{2} 
\int_{z'}^{z} dz'\epsilon_{AB}\gamma^{A}B^{B}\right)\frac{1}{2}
\epsilon\left( z - z'\right)
\label{Omega}
\EEQ
All symbols of integration over $z$ have been
omitted. Eq.~\eq{Hamiltonian} is local in
$t$ and $x^{A}$; those arguments have been suppressed.

The Hamiltonian appearing in \eq{Hamiltonian} is  {\em exact\/.}
However, it is given by a rather complicated, non local  and
non linear expression:
this is the cost we have to pay for explicitly eliminating the 
constraint. We now argue that one can introduce physically reasonable
simplifications, as a result of which the problem becomes a manageable
one. First of all, we notice that the exponential appearing in \eq{Omega}
is of modulus one. Furthermore, it is recognized as the matrix element of an 
operator which is unitary and Hermitean at the same time, hence its
eigenvalues are $\pm 1$. Hence, one expects that at large values 
of $|z - z'|$ the exponent
oscillates rapidly and thus contributes little to the Hamiltonian.
The dominant contribution is thus coming from small values of
 the difference of longitudinal coordinates. Hence, it is reasonable
to approximate the exponential in \eq{Omega} by 1.
In the remaining expression, one term is local in all variables and
the remaining ones are
proportional to $\epsilon\left( z- z'\right)$. Hence, in a Fourier
representation, \VIZ upon writing
\BEQ \phi \left(t, z, x^{A}\right) = \int dk \varphi\left( t, k, x^{A}
\right) \exp \left( -i kz\right)\EEQ
and
\BEQ \epsilon (z) = \frac{{\cal P}}{2\pi i}\int \frac{dk}{k} \exp (-i
kz), 
\EEQ
(In the last equation
${\cal P}$ stands for the principal value.)
Hence,  at high energies ($k\gg m$) 
the Hamiltonian can be approximated by the local term.

Neglecting terms of $O\left( k^{-1}\right)$, the equation of motion for
the density matrix in coordinate representation  reads:
\begin{eqnarray}
-i \partial_{t}\bra{z,\vec{x}}\rho (t)\ket{z', \vec{x'}}& = &\mu \sqrt{2}
\vec{\sigma }\cdot\vec{B}\left(\vec{x},\frac{z-t}{\sqrt{2}}\right)\bra{z,\vec{x}}
\rho (t) \ket{z', \vec{x'}} \nonumber \\
 & - & \mu \sqrt{2}\bra{z ,\vec{x}}\rho
\left(t\right)\ket{z',\vec{x'}}\vec{\sigma}\vec{B}\left(\vec{x'},
\frac{z'-t}{\sqrt{2}}\right)
\label{eqofmotion}
\end{eqnarray}

In this equation, $\vec{x}$ stands for the transverse part of the
coordinate and $ \vec{\sigma}\cdot \vec{B}$ is the  two dimensional
scalar
product in transverse space. Of course,  the coordinate $x^{3}$ had to
be expressed by $ z$ and $t$; hence the $t$-dependence in the magnetic
field.

We choose the initial condition so as to describe a neutrino produced
at $\vec{x}=0$ and with a fixed value of $k$:
\BEQ
\bra{z, \vec{x}}\rho \left(0\right)\ket{z'\vec{x'}}
= \delta^{2}\left( \vec{x}\right)\delta^{2}\left( \vec{x'}\right)
\frac{\exp ik\left(z - z'\right)}{2\pi k}\rho_{s}\left(0\right),
\label{initialcondition} 
\EEQ
where $\rho_{s}(0)$ is the initial value of the spin density matrix.

The variable $k$ being large, the function
$\exp ik\left(z - z'\right)$ is  rapidly oscillating  unless
$z\approx z'$. Therefore, it is permissible to put $z=z'$ in the
coefficient  of the exponential in \eq{initialcondition}. Further, in
the
approximation used,  the
dynamics described by eq.~\eq{eqofmotion}
is independent of $k$ and of $\vec{x}$.
Therefore, the dependence of $\rho (t)$ on $k$ and $\vec{x}$ is
entirely
determined by the initial
condition. Thus, the dynamical equation reduces to an
equation 
involving the spin density matrix alone,  as in non relativistic
spin dynamics. Thus, from now on, we  omit the the subscript $s$ and
we have:
\BEQ
-i\partial_{t}\rho \left( t \right) = \mu \sqrt{2}\left[ \vec{\sigma}\cdot
\vec{B}\left(\frac{z-t}{\sqrt{2}}\right), \rho \left( t\right)\right]
\label{spindynamics}
\EEQ
(Here and in what follows, $\vec{x}=0$ is understood.)

This equation can be solved by the standard time ordered series, \VIZ
\begin{eqnarray}
\rho\left(t\right)&=&\rho\left( 0\right)\nonumber \\
& +&i\mu \sqrt{2} \int_{0}^{t}dt'\left[\vec{\sigma}\cdot
\vec{B}\left(\frac{z-t'}{\sqrt{2}}\right)
,
\rho\left(0\right)
\right]\nonumber \\
 &+& \frac{\left( i\mu \sqrt{2}\right)^{2}}{2!}\int_{0}^{t}
dt'dt''T\left( \left[ \vec{\sigma}\cdot
\vec{B}\left(\frac{z-t'}{\sqrt{2}}\right),
\left[\vec{\sigma}\cdot \vec{B}\left(\frac{z-t''}{\sqrt{2}}\right),
\rho\left( 0 \right)\right]\right]\right)\nonumber \\ &+& \cdots
\label{timeordered}
\end{eqnarray}
We choose the initial condition as:
\BEQ \rho\left( 0\right) = \frac{1}{2}\left( 1 + S\sigma_{3}\right)
, \qquad \left(S^{2}\leq 1\right),
\EEQ
since neutrinos are produced with a definite helicity. (In  the case
of Dirac neutrinos, $S=\pm 1$, depending on whether a neutrino or
anti neutrino
is produced. In the case of Majorana neutrinos, $S$ may assume any
value between the limits stated above, depending on the production
mechanism.)

Next, we average the solution, \eq{timeordered} over the magnetic
field. We choose the generating functional of the moments as follows:
\begin{eqnarray}
Z[j] & = & \int {\cal D}B \exp -\left[ \frac{1}{2}\int
d^{3}xd^{3}x'B_{i}\left(x\right)C^{-1}_{ij}\left(x,x'\right)B_{j}
\left(x'\right)\right]\nonumber \\
& \times & \exp \int d^{3}x j_{i}
\left(x\right)B_{i}\left( x\right );\nonumber \\[1mm]
C_{ij}^{-1} & = & \frac{L}{4\pi \langle B^{2}\rangle}
\left(\delta_{ij}-\frac{\partial_{i}\partial_{j}}{\bigtriangledown^{2}}\right)
\left( L^{-2} - \bigtriangledown^{2}\right)^{2}\delta^{3}\left(x - x'\right).
\label{generatingfunctional}
\end{eqnarray} 
In the last equation, $L$ and $\langle B^{2}\rangle $ stand for the
correlation
length and mean square magnetic field, respectively. The measure is
normalized such that $Z\left[0\right]=1$.  
The transverse
projector is needed in order to make the correlation functions 
solenoidal. With the choice of the tensor $C^{-1}$ given in 
\eq{generatingfunctional}, the leading term in the long distance
behavior of the correlation function is $\propto \exp - \abs{x-x'}$.
In order to average equation \eq{timeordered} over the magnetic field,
one integrates over $B_{3}$ and sets the third component
of the source equal to zero. The  transverse generating functional
reads:
\begin{eqnarray}
Z_{T}&  = &\int {\cal D}\vec{B} \exp - \frac{L}{8\pi \langle B^{2}
\rangle } \int d^{3}x \left[ B^{A}(x) \left( \delta_{AB} -
\frac{\partial_{A}\partial_{B}\left(x\right)}{\vec{\bigtriangledown}^{2}}\right)
B^{B}\right]\nonumber \\ 
& \times & \exp i \int d^{3} x \vec{j}\left(x\right)\cdot \vec{B}\left(x\right)
\label{transversegenerator}
\end{eqnarray}
  
We now notice that in the equation \eq{timeordered}, terms containing
odd
powers of $\mu$ are also odd in $B^{A}$. Therefore, in the limit
$\vec{j} \rightarrow 0$ the average of those terms vanishes. The even
terms in the series are obtained by taking the appropriate functional
derivatives of \eq{transversegenerator}. All of them are expressed in
terms of multiple time integrals of $C_{ij}\left(\abs{t-t'}\right)$
and its powers: those integrations are easily performed. It is
sufficient to illustrate the procedure for the second order term in 
\eq{timeordered}.

Carrying out the integrations, one gets:
\[
 - \mu^{2} \frac{1}{2} S\langle \int_{0}^{t} dt' dt'' \left[
\vec{\sigma}
\cdot \vec{B},\left[
\vec{\sigma}\cdot \vec{B}, \sigma_{3}\right]\right]\rangle
= - \mu^{2}\sigma_{3} \langle B^{2}\rangle tL \left( 1 - \exp
-\frac{t}{L}
\right)
\]
For large times the result in the last equation is
just proportional to $t$. The higher order terms follow a similar
pattern. The end result is:
\BEQ
\langle \rho \left( t \right) \rangle \sim
\frac{1}{2}\left( 1 + S\sigma_{3}\exp - \frac{t}{T}\right),
\label{asymptote}
\EEQ
with
\[ \frac{1}{T} = 2\mu^{2} \langle B^{2}\rangle L . \]

Thus we arrive at the remarkable result that in the random field
the behavior of the helicities is an {\em ergodic} one: irrespective
of what the initial density matrix was, for $t\gg T$, the helicities
are equally distributed. In the case of Dirac neutrinos, this is 
rather uninteresting: roughly 1/2 of them is a sterile one. 
However, calculated neutrino fluxes emerging from such astrophysical objects
as an AGN usually cannot be trusted to an accuracy which would
permit an observational testing of the result. The presence of
sterile Dirac neutrinos does, however, play a role in the early universe,
see~\cite{elmforsetal}; this paper contains a virtually complete
bibliography on the subject\footnote{The authors of those references
typically use an averaging procedure at the level of the evolution
equations, not at the level of solutions. It has been known for some time
that this is not a reliable procedure. For a lucid exposition, see
\cite{vankampen}.}.
  
By contrast, in the case of 
Majorana neutrinos, the two helicity states correspond to 
two different flavors. Given the fact that the neutrinos produced
arise mostly from pion decay, the presence of the other flavor in
roughly equal proportion is an observationally testable result.
We conjecture that the situation is similar if all neutrino flavors
are properly taken into account. 

This work was done during the authors' visit at the Dipartimento
di Fisica, Universit\'{a} di Firenze. We wish to thank Roberto
Casalbuoni, Director of the Department for the hospitality extended to
us. We also thank Bianca Monteleoni for useful conversations on
observational neutrino astrophysics and to Kari Enqvist for
some critical remarks.

\end{document}